\documentclass[11pt]{article}

\usepackage{epsfig}
\usepackage{amsmath}
\usepackage{natbib}

\begin{document}

\title{A Near-Infrared Spectroscopic Survey at the SDSS 2.5-meter Telescope?}

\author{Mike Skrutskie \& John Wilson
\\Astronomy Dept
\\University of Virginia
\\530 McCormick Rd
\\Charlottesville, VA  22903}

\date{August 29, 2005}

\maketitle

\section{Introduction}
Although the world is flush with $> 100$ source/field visible
wavelength spectrographs, no near-infrared equivalent exists.
SDSS possesses a powerful infrastructure for multi-object
spectroscopy which could be adapted for near-infrared observation.
This configuration could, for example, enable dedicated surveys of
the kinematics and chemical evolution of the Milky Way using a
spectrograph with ${\rm R} \sim 3000$ and ${\rm R} \sim 24000$
modes.

\section{Why not a deeper near-infrared imaging survey?}

Wide-field infrared imaging is an alternative attractive future
application for the SDSS telescope.  This alternative should be
considered in context with past and planned large-scale
near-infrared imaging surveys.  2MASS, for example, obtained 100\%
sky coverage to $K < 15$.  It detected effectively every late-type
giant in the Milky Way, detected thousands of L-dwarfs and
$\sim100$ T-dwarfs, and delineated large-scale structure in 4$\pi$
steradians to $z \ge 0.05$.

A significantly deeper near-infrared survey than 2MASS is
scientifically attractive.  For example in 2MASS stars outnumber
galaxies $100:1$ while in a survey three magnitudes deeper than
2MASS the ratio reverses.  A JHK survey to $K \sim 18-19$ would
complement the SDSS depth for galaxy detection and allow
mass-weighted field and cluster luminosity functions, study of
large scale structure to $z \sim 0.2$, and galaxy cluster counts
at $z \sim 0.8-1.0$.  If this survey were to encompass thousands
of square degrees it would increase the volume for L/T dwarf
detection by 1000 fold and allow detection of $T_{\rm eff} = 400
{\rm K}$ brown dwarfs within one parsec.  Also in the rare object
venue, a $K \sim 18-19$ survey would enable $z > 6$ quasar counts
using SDSS $z'$ band dropouts.

Dedicating more time to smaller areas enable surveys of square
degree regions to $K\sim 21-23$. Such a moderately deep
pencil-beam survey would enable observation of galaxy evolution in
the rest-frame visible at $z=3$.

Two deep near-infrared imaging surveys are underway, however:
UKIRT Infrared Deep Sky Survey (UKIDSS; {\tt
http://www.ukidss.org/}) and Visible and Infrared Survey Telescope
for Astronomy (VISTA; {\tt http://www.vista.ac.uk/}). UKIDSS is in
progress on the UKIRT $3.5\,{\rm m}$ telescope equipped with a
0.21 sq. degree imager.  This survey expects to observe 7200 sq.
degrees to $K=18.5$ in $ZYJHK$ bands, 35 sq. degrees to $K=21$,
and 1 sq. degree to $K=23$. VISTA expects to start this year on a
12 year survey using a dedicated $4\,{\rm m}$ telescope in Chile
with a 1 sq. degree imager. This survey expects to observe 5000
sq. degrees in $JHK$ to $K=20.5$ and 25 sq. degrees to $K=22.5$.

These other surveys, despite completion dates well into the
future, are nevertheless well-established and ripe to deliver
results.  An SDSS near-infrared imaging survey would arrive late
and be a small player looking for niche science.

\section{A near-infrared spectroscopic alternative?}

The SDSS $2.5\,{\rm m}$ could become a near-infrared spectroscopic
survey telescope delivering spectra for $\sim 1000$ sources per
night. Anticipated sensitivities are $H \sim 14$ at $R=3000$ and
$H \sim 12$ at $R=20000$ for ${\rm SNR}=10$ in 30 minutes.  Every
potential target for such a spectrographic survey has thus been
detected at high SNR by 2MASS in $JHK$ which enables good source
selection via color discrimination.  Cool giant stars in the Milky
Way are natural targets, both because they are intrinsically
luminous at near-infrared wavelengths and because galactic
extinction is significantly smaller at these wavelengths.  Spectra
in this wavelength regime can provide classification, kinematics
and abundance estimates.

\section{Practical Issues}

The most straightforward implementation of near-infrared
spectroscopy would use the SDSS fiber and plug plate
infrastructure as is.  The near-infrared opacity of both the
fibers and common corrector could be limiting factors depending on
the scientifically driven bandpass selection.

\begin{figure}[h]
   \begin{center}
   \begin{tabular}{c}
   \includegraphics[scale=0.5]{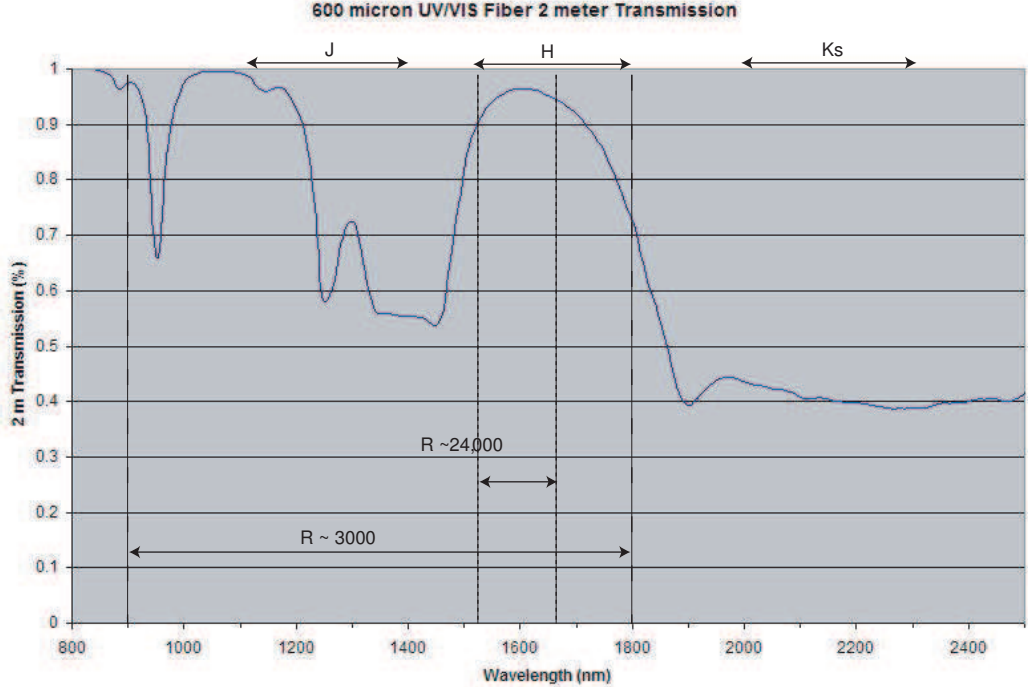}
   \end{tabular}
   \end{center}
   \caption[example]
%>>>> use \label inside caption to get Fig. number with \ref{}
   { \label{fig:attenuation}
Transmission over $2\,{\rm m}$ path length of $600\,\mu {\rm m}$
high-OH fiber manufactured by Polymicro.  The data is scaled from
$10\,{\rm m}$ transmission measurements provided by Ocean Optics,
Inc.  The 2MASS $JHK$ photometric bands and wavelength coverage
provided by a strawman spectrograph (described in Section
\ref{sec:strawman}) are identified.}
   \end{figure}

\subsection{Fiber Transmission}
The current SDSS fibers have high-OH content (so-called `wet
fibers') to improve UV performance.  OH absorption is significant
in some portions of the near-infrared.  The SDSS fiber is
$180\,\mu {\rm m}$ single strand UV-enhanced (high-OH) step-index
fiber.\footnote{The fiber, known as `FV' fiber, was purchased from
Polymicro. Polymicro no longer manufactures this product; its
current UV-enhanced fiber is known as `FHP'.}  Figure
\ref{fig:attenuation} shows transmission of modern Polymicro
high-OH fiber in the near-infrared over a $2\,{\rm m}$
path-length. Over half of the long wavelength side of $J$ band is
adversely affected by the fiber transmission, with in-band
transmission as low as $55\%$ due to OH absorption.  The existing
fibers have reasonably good transmission over much of the H-band,
peaking at about $95\%$ at $1.6\,\mu {\rm m}$ in the center of the
H-band.  Science which exploits the H-band can proceed without
fiber replacement.  Beyond H-band transmission falls to a nearly
constant $40\%$ transmission beyond $1.85\,\mu {\rm m}$ and
through all of $K$ band.

\begin{figure}[h]
   \begin{center}
   \begin{tabular}{c}
   \includegraphics[scale=1.25]{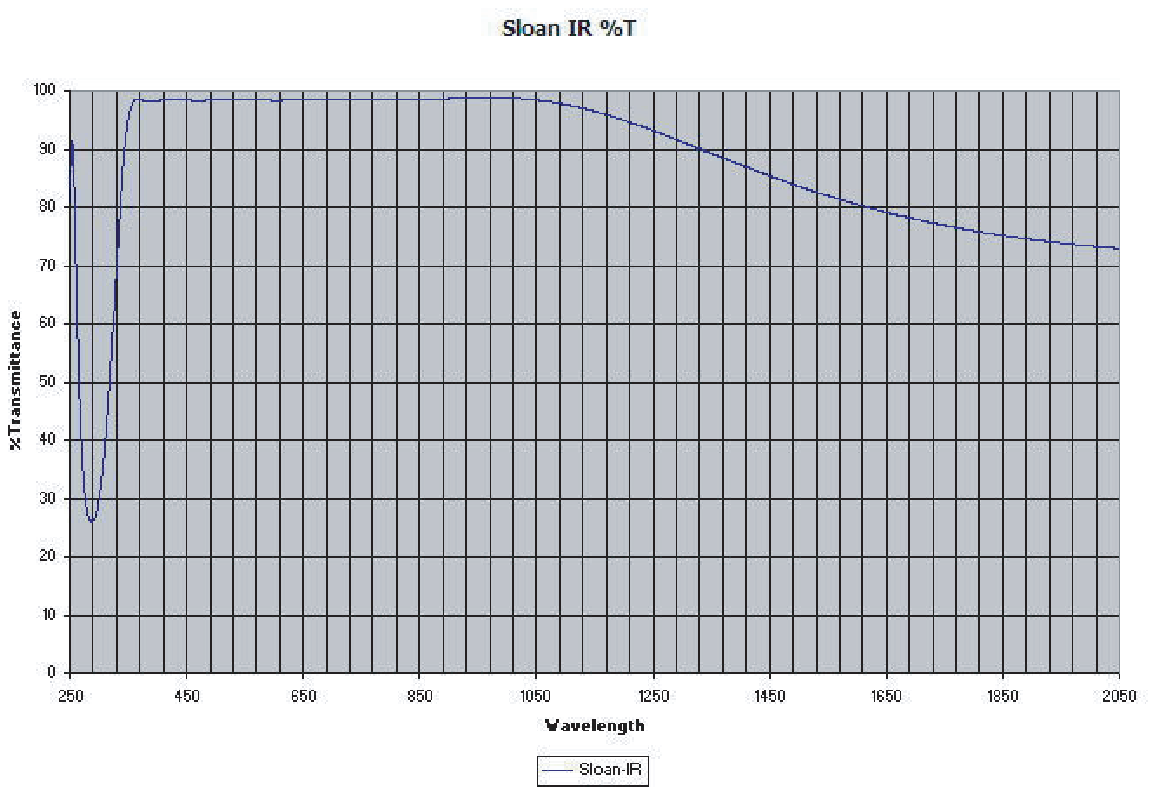}
   \end{tabular}
   \end{center}
   \caption[example]
%>>>> use \label inside caption to get Fig. number with \ref{}
   { \label{fig:coating}
Design visual and NIR transmission through the coated fused silica
common corrector (includes transmission through two surfaces plus
bulk material transmission) located within the central hole of the
primary mirror. Data from Infinite Optics, Inc., original coating
vendor.}
   \end{figure}

\subsection{Corrector Plate Transmission}
The Common Corrector inserted within the hole of the primary
mirror is made from Corning 7940 \footnote{Corning no longer
manufactures this material. Its current equivalent is 7980.} Fused
Silica.  The raw material transmission is high shortward of
$2.0\,\mu {\rm m}$, but starts to degrade longward of this
wavelength.  The corrector coating\footnote{QSP Optics did the
coating; this company is now called Infinite Optics.} is optimized
for $0.32-1.1\,\mu {\rm m}$.  The design visual and near-infrared
transmission curves from the coating vendor through both coated
surfaces and the lens material is shown in Figure
\ref{fig:coating}. The corrector could be replaced with modest
cost and effort if necessary or recoated to minimize reflectivity
over a broader bandpass.   Such a coating would likely come at the
cost of a slight compromise of current performance at visible
wavelengths.

\subsection{Thermal background}
A $1.6\,\mu {\rm m}$ spectrograph could be a room temperature
instrument, or one which is modestly refrigerated in a dessicated
atmosphere. Beyond $1.6\,\mu {\rm m}$, ambient thermal background
begins to compromise spectrographic sensitivity considerably,
particularly for a fiber configuration where the spectrograph has
to view the warm fiber ends and their holders.  Although K-band is
rich in scientific opportunity, sensitivity lost to thermal
background, transmission issues, and the overall increase in
complexity of the resulting instrument may lessen interest in this
wavelength regime.

\subsection{Dynamic Range Considerations}
Near-infrared array multiplexers, unlike CCDs, are capable of
reading out non-destructively many times per exposure (because the
collected charge stays local to pixels and each pixel is addressed
in turn by the multiplexer).  A single long exposure can be broken
into many full-frame sub-exposures of different lengths enabling
the extraction of bright source spectra, or bright spectral
regions or lines, prior to reaching saturation while preserving a
deep integration on the array as a whole. A huge range of source
flux can be addressed in a single plug plate.

\section{Specific science suggestion: A Galactic Kinematic and Chemical
Evolution Survey}

\begin{figure}[h]
   \begin{center}
   \begin{tabular}{c}
   \includegraphics[scale=0.2]{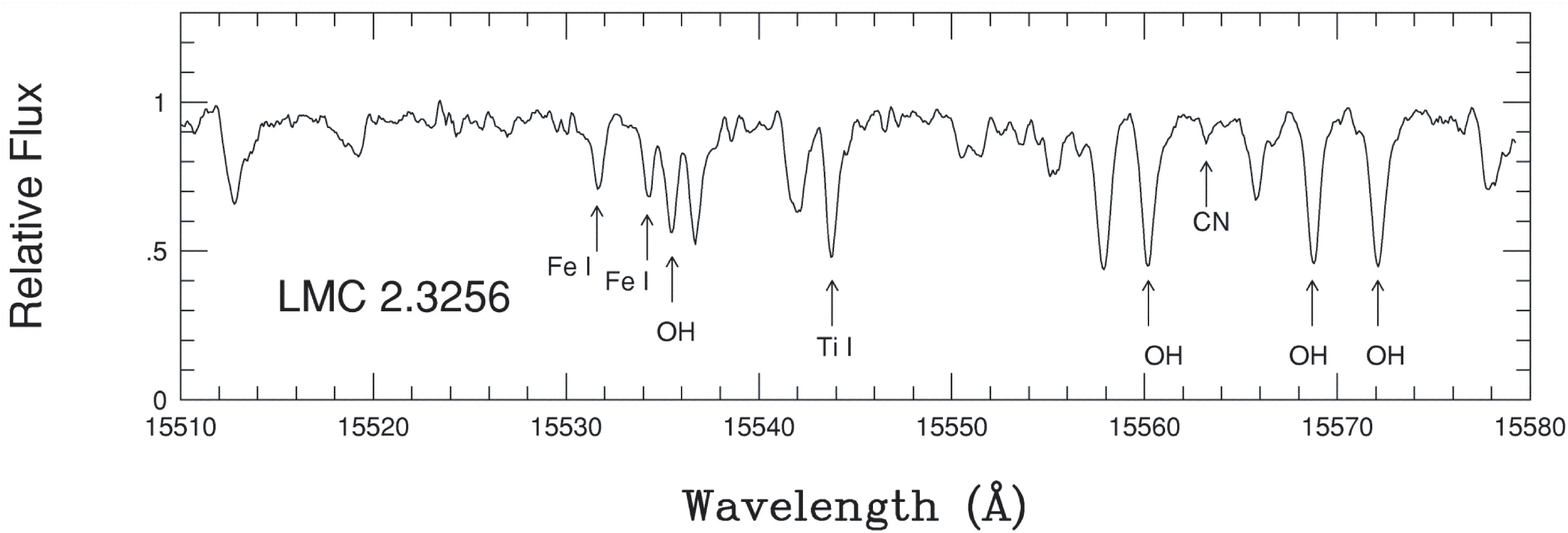}
   \end{tabular}
   \end{center}
   \caption[example]
%>>>> use \label inside caption to get Fig. number with \ref{}
   { \label{fig:smith-spectra}
Molecular CNO lines in an LMC red giant \citep{sea02}.}
   \end{figure}

Carbon, Nitrogen and Oxygen are tracers of Galactic star formation
history and are sensitive (relative to Fe) to the ratio of Type II
to Type Ia SN (see e.g. \citet{sea02}). CNO have weak atomic lines
and are often observed under non-LTE conditions in the visible,
making abundance measurements there difficult.  Molecular CNO
lines, e.g. CN, CO and OH, are more reliable abundance diagnostics
and are effective when observed with $R > 20000$.  A critical
spectral region lies between 1.5 and $1.7\,\mu {\rm m}$ in the $H$
band (Figure \ref{fig:smith-spectra}) - just the range of
wavelengths where the existing fiber transmission is good.  This
wavelength range is rich in diagnostic atomic and molecular
absorption lines.  High resolution near-infrared spectroscopy has
been exploited by atleast two groups to date, including
\citet{org02,org05} abundance studies of globular clusters and
\citet{sea02} abundance studies of LMC red-giants. SDSS-NIR has
the potential to extend this work to the broader galactic
population combining metalicity information with the precise
kinematic measurements which naturally derive from observations at
high spectral resolution.  Such a survey would be dominated by
targets in or near the Galactic plane (leveraging the extinction
advantage of working in the near-infrared). A color selection from
2MASS yields 11 million potential giant star targets with $H < 12$
visible from Apache Point. This plane survey could operate in
tandem with a complementary high galactic latitude program which
focuses on both infrared and visible wavelength targets to fill
out a fiber plate -  assuming only one of the two existing
spectrographs are replaced.

\section{An attractive array technology}

If the primary science can be addressed at wavelengths shortward
of $1.7\,\mu {\rm m}$, InGaAs detector/array technology becomes an
attractive and potentially cost effective alternative.  The long
wavelength cut-off of standard InGaAs material is $1.7\,\mu {\rm
m}$. Arrays of these detectors can be manufactured less
expensively than HgCdTe and InSb and are becoming competitive in
dark current and noise performance.

UVA collaborates with Sensors Unlimited, Inc. in the development
of astronomical InGaAs technology.  
Recently InGaAs arrays were successfully bonded to low-noise
astronomical 1K x 1K multiplexers (for SNAP).  2K x 2K arrays are
on the horizon.  One can anticipate substantial progress in the
development of InGaAs arrays with low dark current and good read
noise characteristics in time for implementation in an infrared
spectrograph for the 2.5-meter.

\section{Strawman instrument considerations\label{sec:strawman}}

One could construct a spectrograph akin to the SDSS optical
spectrograph which initially uses the existing fiber and plug
plates. Should the spectral coverage demanded by science
considerations make it necessary, the existing fibers could be
replaced with low-OH counterparts in a mechanical configuration
identical to the current fiber configuration (although replacement
of fibers also allows consideration of constructing a more stable
floor-standing spectrograph). The spectrograph would contain two
interchangeable gratings to provide two modes: an $R \sim 3000$
``faint-object'' mode for classification and kinematics ($\sim
5\,{\rm km\,s^{-1}}$) to $H=14$ in 30 minutes covering
$0.9-1.8\,\mu {\rm m}$ and an $R \sim 24000$ mode for
classification, kinematics ($\sim 1\,{\rm km\,s^{-1}}$) and
abundances to $H=12$ in 30 minutes covering $1.52-1.66\,\mu {\rm
m}$.  The high resolution grating could be rotatable to allow
selection of the wavelength coverage within $H$ band.

This spectrograph would require 4096 pixels in the dispersion
direction and thus a detector format similar to the existing SDSS
optical spectrographs.  Thus two 2K x 2K InGaAs arrays would be
adequate. This array dimension would permit $> 200$ spectra per
exposure.

\section{Some initial technical questions}

\begin{itemize}
\item The transmission estimates reported here should be verified
by
 empirical tests with the actual fiber material (ultimately with
 a total throughput test at the telescope).

\item The current fibers are sized at 3'' as a compromise between
  point source and extended source sensitivity for SDSS.
  Fiber replacement would provide an opportunity to revisit
  the fiber spatial scale.

\item Is it feasible to replace a portion of th existing fibers
 with smaller core low-OH fibers to devote to fainter sources or
 or to assemble into IFU `dense pak' configurations.

\item What are the size and flexure requirements of an $R \sim
24000$ spectrograph?  Is it feasible to mount it on the back of
the SDSS telescope, and thus re-use the existing fiber
infrastructure ($2\,{\rm m}$ fibers).  Details to consider:
grating type/size, camera focal length, sampling requirements,
wavelength coverage per object, number and type of detectors.

\item Should the instrument be designed to allow use at the
$3.5\,{\rm m}$ telescope as well?

\end{itemize}


\begin{thebibliography}{}

\bibitem[Smith et al.(2002)]{sea02}Smith, V. V. et al., 2002, $AJ$, \textbf{124}, 3241
\bibitem[Melendez and Barbuy(2002)]{mb02}Melendez, J. and Barbuy, B.,
  2002, {\it Ap. J.}, \textbf{575}, 474.
\bibitem[Origlia et al.(2005)]{org05}Origlia, L., Valenti, E. and
  Rich, R. M., {\it M.N.R.A.S.}, \textbf{356}, 1276.
\bibitem[Origlia et al.(2002)]{org02}Origlia, L., Rich, R.M., and
Castro, S., 2002, {\it  Astron. J.}, \textbf{123}, 1559.

\end{thebibliography}
\end{document}